\documentclass[a4paper]{jpconf}
\usepackage{graphicx}
\usepackage{psfrag}
\begin{document}
\title{Towards a closed differential aging formula in  special
relativity}

\author{E Minguzzi}

\address{Department of Applied Mathematics, Florence
 University, Via S. Marta 3, I-50139 Florence, Italy.}

\ead{ettore.minguzzi@unifi.it}

\begin{abstract}

It is well known that the Lorentzian length of a timelike curve in
Minkowski spacetime  is smaller than the Lorentzian length of the
geodesic connecting its initial and final endpoints. The difference
is known as the {\em differential aging} and its calculation in
terms of the proper acceleration history of the timelike curve would
provide an important tool for the autonomous spacetime navigation of
non-inertial observers. I give a solution in 3+1 dimensions which
holds whenever the acceleration is decomposed  with respect to a
lightlike transported frame (lightlike transport will be defined),
the analogous and more natural problem for a Fermi-Walker
decomposition being still open.
\end{abstract}

\section{Introduction}
In this work I consider the problem of determining the spacetime
position, with respect to an inertial frame $K$, of an arbitrarily
moving observer $O$ in Minkowski spacetime $M$. The restriction to
Minkowski spacetime  means that we are considering travellers $O$
which move on a region of spacetime which can be considered
approximately flat.

Usually this problem is solved by assuming to be known the velocity
of the non-inertial observer $O$, ${\bf v}(t)$, with respect to the
inertial frame time $t$. In this way also his position ${\bf x}(t)$
is known and hence his spacetime position $(t,{\bf x}(t))$
parametrized with respect to $t$. The non-inertial observer asking
his position at proper time $\tau$ may obtain it by direct
communication with the inertial observer. Indeed, through the
inversion of the relation
\begin{equation}
\tau(t)=\tau_0+\int_{t_0}^{t}\sqrt{1-{\bf v}^2(t)}\, {\rm d} t ,
\end{equation}
the inertial observer can obtain his spacetime position as a
function of proper time.

However, this procedure has some important drawbacks. For instance,
since the needed observables ({\bf v}(t)) are measured by the
inertial observer, it requires a communication between the inertial
and the non-inertial observers which becomes more and more unlikely
as the distance between them increases.

Fortunately, the non-inertial observer may obtain the spacetime
position without referring to external observers through measures
performed only in the local comoving laboratory. The idea is to use
the acceleration history of the observer and to reconstruct the
timelike trajectory on spacetime from that information.

Thus, imagine the non-inertial observer transporting an
accelerometer and $n$ mutually orthogonal gyroscopes which together
with the covariant velocity define a Fermi-Walker transported frame
(the spacetime has $n+1$ dimensions). Through the accelerometer he
can measure his own acceleration in intensity and direction, in
particular he can read the components $\tilde{a}^i(\tau)$,
$i=1,\ldots,n$, with respect to the directions defined by the
gyroscopes. The problem is that of recovering the spacetime position
$x^{\mu}(\tau)$ from the historical proper acceleration data
$\tilde{a}^i(\tau)$. Since the inertial frame is not fixed by the
data $\{\tilde{a}^i(\tau)\}$ the answer will be unique only up to a
Poincar\'e transformation. In particular we may fix the origin of
the inertial frame so that it coincides with the event $\tau=0$ on
$O$'s curve $x(\tau)$, and the final answer $x^{\mu}(\tau)$ will be
determined only up to a Lorentz transformation.

In this work the timelike convention is used, $\eta_{00}=1$, and
units are such that $c=1$. In boldface we denote $n$-vectors and the
scalar product between boldfaced vectors is the usual Euclidean one.
\begin{figure}[!ht]
\centering \psfrag{T}{$T$} \psfrag{X}{$x(\tau)$}
\includegraphics[width=4cm]{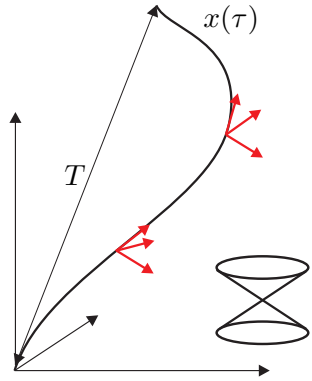}
\caption{Fermi-Walker transport in 2+1 dimensional Minkowski
spacetime. The problem is to determine the spacetime position in
terms of the components of the proper acceleration. A first step is
to calculate $T^{2}(\tau)=x^{\mu}(\tau)\eta_{\mu
\nu}x^{\nu}(\tau)$.} \label{fig1}
\end{figure}

\section{The differential aging}
Fix a value of $\tau>0$ and consider the timelike geodesic $\eta$
connecting $x(0)$ with $x(\tau)$. It can be interpreted as the
worldline of an inertial observer $\bar{O}$ which leaves $O$ at time
$\bar{\tau}=\tau=0$ and meets again $O$ at time $\bar{\tau}=T(\tau)$
where $T$ is given by $T^{2}(\tau)=x^{\mu}(\tau)\eta_{\mu
\nu}x^{\nu}(\tau)$. In the usual twin examples, $O$ and $\bar{O}$
are twins and $\Delta(\tau)=T(\tau)-\tau>0$ gives the difference in
aging at the second meeting event $x(\tau)$.

The calculation of $T(\tau)$ and hence of the differential aging in
terms of the acceleration would represent an important first step in
the solution of the navigation problem. Moreover, it would provide a
sort of {\em inertial clock} $T(\tau)$ which would give to the
non-inertial observer the information about his difference in age
with respect to an imaginary twin, all this without any use of the
concept of distant simultaneity (see the discussion in
\cite{minguzzi04c}). In particular, by varying $\tau$ we are
actually considering the differences in age with respect to a
1-parameter family of imaginary free falling twins. This problem has
been solved at least in 1+1 dimensions \cite{minguzzi04c}, the
solution being given by
\begin{equation} \label{onl}
\label{itf} T^{2}(\tau)=\Big[\!\int^{\tau}_{0} e^{\int^{\tau'}_{0}
\tilde{a}(\tau'\!')d \tau'\!'} d \tau'\Big] \,\Big[\!\int^{\tau}_{0}
e^{-\int^{\tau'}_{0} \tilde{a}(\tau'\!')d \tau'\!'} \, d \tau'\Big].
\end{equation}
where $\tilde{a}$ is the only component of the acceleration as
measured by a coming accelerometer. In this work I will obtain a
generalization to n+1 dimensions.

\section{The navigation problem}

Let us find the equations of spacetime navigation in terms of the
acceleration.

The definition of covariant ($n+1$)-velocity and covariant
($n+1$)-acceleration is ($\mu=0,\ldots,n$)
\begin{eqnarray}
\frac{{\rm d} x^{\mu}}{{\rm d} \tau} &=& u^{\mu} ,\\
\frac{{\rm d} u^{\mu}}{{\rm d} \tau} &=& a^{\mu},
\end{eqnarray}
where to be rigorous the velocity and acceleration are the vectors
$u^{\mu} \partial/\partial x^{\mu}$, $a^{\mu}\partial/\partial
x^{\mu}$, i.e. the above $a^{\mu}$, $u^{\mu}$, are just the
components with respect to the tetrad $\partial/\partial x^{\mu}$ of
the inertial observer at rest in $K$. The non-inertial observer $O$
has a different tetrad ${e}_{\mu}$ which we can identify with its
4-velocity ${e}_{0}=u^{\mu} \partial/\partial x^{\mu}$, and the
directions of $n$ gyroscopes ${e}_i$. These last directions can be
decomposed with respect to $\partial/\partial x^{\mu}$
\begin{equation}
e_{i}=e_{i}^{\mu} \frac{\partial}{\partial x^{\mu}}.
\end{equation}
The components $\tilde{a}^{i}(\tau)$, of the acceleration with
respect to the gyroscopes $e_i$, are given, so that
\begin{equation}
a^{\mu} \frac{\partial }{\partial x^{\mu}}=\tilde{a}^{i}(\tau)\,
{e}_{i}
\end{equation}
or
\begin{equation}
a^{\mu}(\tau)=\tilde{a}^{i}(\tau)\, e^{\mu}_{i}(\tau),
\end{equation}
 and our task is the calculation of $T=\sqrt{x^{\mu}(\tau)\eta_{\mu
\nu} x^{\nu}(\tau)}$, under the assumption $x^{\mu}(0)=0$,
$u^{i}(0)=0$, that is, the non-inertial observer starts at rest and
from the origin of the inertial frame $K$. First, we have to find
$e_{i}^{\mu}(\tau)$. The evolution of the tetrad is given by the
equation of the Fermi-Walker transport \cite {misner73}
\begin{equation}
\frac{{\rm d}}{{\rm d} \tau} {e}^{\mu}_i=(a^{\mu}
u_{\nu}-u^{\mu}a_{\nu}) e^{\nu}_{i}.
\end{equation}
Using the orthogonality of the comoving tetrad $\{e_0=u, e_i\}$,
($e_i \cdot e_j=-\delta_{ij}$, $e_i \cdot e_0=0$, $e_0 \cdot e_0=1$)
it reads
\begin{equation}
\frac{{\rm d}}{{\rm d} \tau} {e}^{\mu}_i=u^{\mu}\,
\tilde{a}^{i}(\tau), \qquad i=1,\ldots,n ; \ \mu=0,\ldots,n.
\end{equation}

Summarizing, the navigation problem of recovering the timelike
worldline in terms of the components of the proper acceleration
taken with respect to a Fermi-Walker transported frame takes the
form (I have made explicit the summation over the indexes)
\begin{eqnarray}
\frac{{\rm d} x^{\mu}}{{\rm d} \tau} &=& u^{\mu},  \qquad  \mu=0,\ldots,n, \\
\frac{{\rm d} u^{\mu}}{{\rm d} \tau} &=& \sum_{i=1}^{n} \tilde{a}^{i}(\tau) e^{\mu}_{i}(\tau), \qquad  \mu=0,\ldots,n, \\
\frac{{\rm d}}{{\rm d} \tau} {e}^{\mu}_i&=&u^{\mu}
\tilde{a}^{i}(\tau), \qquad i=1,\ldots,n ; \ \mu=0,\ldots,n.
\end{eqnarray}
In particular if the inertial frame $K$ is chosen such that $O$ is
at rest and oriented as $K$ at time $\tau=0$ then the initial
conditions are $u^{i}(0)=0$, $u^0(0)=1$, $x^{\mu}(0)=0$,
$e^{\mu}_i(0)=\delta^{\mu}_i$. Through these (n+1)(n+2) equations
one finds $x^{\mu}(\tau)$ and hence $T(\tau)$ which represents the
inertial time dilation with respect to an inertial observer that
would see the journey of $O$ till time $\tau$ as closed.

No closed analytical  solution of these equations is known,
nevertheless the problem of obtaining an analytical solution for
$T(\tau)$ may admit a solution. The reason is that $T$ being a
Lorentz invariant is independent of the initial conditions and the
problem may therefore simplify. The following section brings
evidence in favor of this conclusion.

\section{A closed formula for the differential aging}
The previous system of differential equations can be decomposed on
$n+1$ systems, one for each value of $\mu$. In order to find
$T^2(\tau)=x^{\mu} x_{\mu}$ we define
\begin{equation}
y_{0}=x_{\mu} e^{\mu}_{0}; \qquad y_{i}=x_{\mu} e^{\mu}_{i},
\end{equation}
so that
\begin{equation}
T^2=y_{\alpha}y^{\alpha}=y_0^2-{\bf y}^2.
\end{equation}
The reader may convince him or herself that $y^{\alpha}$ is the
difference of the inertial coordinates of $x(\tau)$ and $x(0)$ with
respect to an inertial frame centered at $x(\tau)$ and oriented as
the frame $e_i(\tau)$.

From the previous system of differential equations we obtain
\begin{eqnarray}
\frac{{\rm d} T^2}{{\rm d} \tau}&=&2 y_0, \label{pri}\\
\frac{{\rm d} y_0}{{\rm d} \tau}&=& 1+{\bf \tilde{a}} \cdot {\bf y}, \label{sec} \\
\frac{{\rm d} {\bf y}}{{\rm d} \tau}&=& y_0 {\bf \tilde{a}},
\label{poi}
\end{eqnarray}
with the initial conditions $y_{\alpha}(0)=0$, $T(0)=0$. The first
equation follows also from the last two.

Define $y=\vert {\bf y}\vert$,  ${\bf \hat{y}}={\bf y}/y$, ($-{\bf
\hat{y}}$ represents the direction of $x(0)$ with respect to the
mentioned frame centered at $x(\tau)$) and multiply Eq. (\ref{poi})
by $2 {\bf y}$
\begin{equation}
\frac{{\rm d} y}{{\rm d} \tau}= y_0 ({\bf \tilde{a}} \cdot {\bf
\hat{y}}),
\end{equation}
to be considered in system with the equation
\begin{equation}
\frac{{\rm d} y_0}{{\rm d} \tau}= 1+y ({\bf \tilde{a}} \cdot {\bf
\hat{y}}).
\end{equation}
Sum and subtract both equations to obtain
\begin{eqnarray}
\frac{{\rm d} (y_0+y)}{{\rm d} \tau}&=& 1+({\bf \tilde{a}} \cdot {\bf \hat{y}})(y_0+y), \\
\frac{{\rm d} {(y_0- y)}}{{\rm d} \tau}&=& 1- ({\bf \tilde{a}} \cdot
{\bf \hat{y}}) (y_0-y).
\end{eqnarray}
Hence multiplying by $\exp(-\int^{\tau}_{0} ({\bf \tilde{a}} \cdot
{\bf \hat{y}}) {\rm d} \tau)$ the former and by
$\exp(\int^{\tau}_{0} ({\bf \tilde{a}} \cdot {\bf \hat{y}}) {\rm d}
\tau)$ the latter
\begin{eqnarray}
\frac{{\rm d} }{{\rm d} \tau}\{ (y_0+y) \exp(-\int^{\tau}_{0} ({\bf
\tilde{a}}
 \cdot {\bf \hat{y}}) {\rm d} \tau)\}&=& \exp(-\int^{\tau}_{0} ({\bf \tilde{a}} \cdot {\bf \hat{y}}) {\rm d} \tau), \\
\frac{{\rm d} {}}{{\rm d} \tau}\{(y_0- y) \exp(\int^{\tau}_{0} ({\bf
\tilde{a}} \cdot {\bf \hat{y}}) {\rm d} \tau)\}&=&
\exp(\int^{\tau}_{0} ({\bf \tilde{a}} \cdot {\bf \hat{y}}) {\rm d}
\tau),
\end{eqnarray}
and finally, using $T^2=(y_0+y) (y_0- y)$,
\begin{equation}
T^2= \{ \int_{0}^{\tau} e^{-\int^{\tau'}_{0}( {\bf \tilde{a}} \cdot
{\bf \hat{y}}) {\rm d} \tau''}{\rm d} \tau' \}\{ \int_{0}^{\tau}
e^{\int^{\tau'}_{0} ({\bf \tilde{a}} \cdot {\bf \hat{y}}) {\rm d}
\tau''}{\rm d} \tau' \} .
\end{equation}
Thus, the differential aging $\Delta(\tau)$ depends only on the
component $({\bf \tilde{a}} \cdot {\bf \hat{y}})$ of the
acceleration.  Note that this formula makes sense only if for all
$\tau'$, $y(\tau')\ne0$, otherwise ${\bf \hat{y}}$ is not defined.
Unfortunately, in $n+1$ dimensions it is not easy to find how the
direction ${\bf \hat{y}}(\tau)$ depends on the acceleration history
so that this formula, while giving some insights, does not solve the
problem.

Its physical meaning can be understood through the following
reasoning. Imagine that at event $x(0)$ an explosion takes place so
that the remnants of this explosion begin moving in all directions
at a constant velocity with respect to the inertial frame $K$. The
free falling observer $\bar{O}$ can be identified with one of this
remnants. Now, the non-inertial observer observes at any time $\tau$
a flow of remnants of direction ${\bf \hat{y}}(\tau)$ passing nearby
his comoving laboratory. The above formula states that in order to
calculate the differential aging, $O$ does not need to record all
the components of the proper acceleration but only the one of
direction ${\bf \hat{y}}(\tau)$, i.e. the one along the remnants'
flow. While interesting this result has limited applicability as we
can not always assume, in practice, that an explosion has taken
place at event $x(0)$ so that the direction of the remnants'
velocity becomes locally observable.

Note that in 1+1 dimensions since there is only one component of the
acceleration $({\bf \tilde{a}} \cdot {\bf \hat{y}})=\pm \tilde{a}$
whenever $y$ does not change sign so that in this domain the formula
reduces to (\ref{onl}).

\section{The parallel transport which keeps a lightlike direction fixed}

A different strategy was followed in \cite{minguzzi05e}. The idea is
to replace the Fermi-Walker transport with a different kind of
transport. It is well known that software guided telescopes keep
their orientation towards a given star under observation despite the
fact that the earth rotates. The orientation of the telescope does
not change according to the Fermi-Walker transport but according to
a transport introduced in \cite{minguzzi05e}, that for short I
termed {\em lightlike parallel transport}. It must not be confused
with the generalization of the Fermi-Walker transport above a
lightlike curve as studied, for instance, in \cite{samuel00,bini06}.
Rather, it is a minimal modification of the Fermi-Walker transport
over a timelike curve which keeps the direction of a covariantly
constant lightlike vector $n$ (interpreted as the light coming from
a star) unchanged with respect to a lightlike transported frame.

A telescope fixed with respect to a lightlike transported frame
would always be pointing towards the same star, moreover this
transport depends on the star (i.e. null direction $n$) chosen.

If $v$ is a vector field over $x(\tau)$, then the lightlike
transport is defined through the equation
\begin{equation}
\nabla_{u}v_{\mu}-\nabla^{L}_{u}v_{\mu}=\Omega^{L}_{\mu \nu}v^{\nu},
\end{equation}
where $\nabla$ is the usual Levi-Civita connection, and where
\begin{eqnarray}
\Omega^{L}_{\mu
\nu}&=&\frac{1}{u^{\beta}n_{\beta}}[a_{\mu}n_{\nu}-a_{\nu}n_{\mu}].
\end{eqnarray}
As a consequence the Fermi-Walker frame and the lightlike
transported frame rotate one with respect to the other but the
relative angular velocity is usually very small. This rotation is
required in order to correct for the effect of aberration of light
that would change the night sky position of the selected star.

The crucial observation is that the lightlike parallel transport can
be obtained in practice by suitably correcting, time by time, the
orientation of the frame so that the selected star stays always in
the same position in the night sky. In fact, that this transport is
feasible is proved by the mentioned examples of telescopes.

Let $e_i^{L}=e_i^{L\, \mu}\partial/\partial x^{\mu}$ be the
lightlike transported frame. The acceleration measured by the
comoving accelerometer can now be projected with respect to the axes
of the lightlike transported frame $a^{\mu}=\check{a}^{i}(\tau)
e_i^{L\, \mu}(\tau)$ and the problem becomes now that of obtaining
the trajectory on spacetime, $x(\tau)$, staring from the data
$\{\check{a}^{i}(\tau) \}$.

Remarkably, this problem can be solved completely, a solution being
given in \cite{minguzzi05e}. Here we give only the solution for
$T(\tau)$. Note that since the position of the selected star does
not change, we can always orient the lightlike frame so that the
light from the star goes on direction $e^{L}_{n}$. The components of
the acceleration can then be decomposed in acceleration along the
direction of light, $\check{a}_{\parallel}$, and acceleration
perpendicular to it, ${\bf \check{a}}_{\perp}$, where this time the
boldface denotes an $(n-1)$-vector. The solution is then
\begin{eqnarray*}
T^2(\tau)&=&  [\int_{0}^{\tau} e^{-\!\int_{0}^{\tau'}\!\!\check{{\rm
a}}_{\parallel} {\rm d} \tau'\!'}{\rm d} \tau'] [\int_{0}^{\tau}
e^{-\!\int_{0}^{\tau'}\!\!\check{{\rm a}}_{\parallel} {\rm d}
\tau'\!'} (\int_{0}^{\tau'} e^{\int_{0}^{\tau'\!'}\!\!\check{{\rm
a}}_{\parallel} {\rm d} \tau'\!'\!'} \check{{\bf a}}_{\perp}
{\rm d} \tau'\!')^2 {\rm d} \tau'] \\
&& \!\!\!  \!\!\!  \!\!\!  \!\!\!  \!\!\!  \!\!\! -[\int_{0}^{\tau}
e^{-\!\int_{0}^{\tau'}\!\!\check{{\rm a}}_{\parallel} {\rm d}
\tau'\!'} (\int_{0}^{\tau'} e^{\int_{0}^{\tau'\!'}\!\!\check{{\rm
a}}_{\parallel} {\rm d} \tau'\!'\!'} \check{{\bf a}}_{\perp} {\rm d}
\tau'\!') {\rm d} \tau']^2 + [\int_{0}^{\tau}
e^{-\!\int_{0}^{\tau'}\!\!\check{{\rm a}}_{\parallel} {\rm d}
\tau'\!'}{\rm d} \tau'][\int_{0}^{\tau}
e^{\int_{0}^{\tau'}\!\!\check{{\rm a}}_{\parallel} {\rm d}
\tau'\!'}{\rm d} \tau'],
\end{eqnarray*}
and using the Cauchy-Schwarz inequality one can show that
$T(\tau)>\tau$ as expected, unless $\check{{\rm a}}_{\parallel}={
0}$ and $\check{{\bf a}}_{\perp}=0$, in which case $T=\tau$.

\section{Conclusions}
I have considered the problem of reconstructing the trajectory of a
non-inertial observer starting from the knowledge of the proper
acceleration. This problem is hard to solve if the data is given in
terms of the components of the acceleration with respect to a
Fermi-Walker transported frame while it can be solved if the newly
introduced {\em lightlike parallel transport} is used. The offered
solution is satisfactory as it can be implemented operationally;
nevertheless, it would be nice to solve  the problem of finding a
closed differential aging formula even in the case of a Fermi-Walker
decomposition of the acceleration. This problem remains still open
and represents an interesting challenge for future investigations.

\section*{Acknowledgments}
I thank J.-F. Pascual-S{\'a}nchez for pointing out references
\cite{samuel00,bini06}. \\


\end{document}